\documentstyle[prl,aps,twocolumn,epsf]{revtex}
\def\break#1{\pagebreak \vspace*{#1}}
\begin{document}
\draft
\title{Electroneutrality and the Friedel sum rule in a Luttinger liquid}
\author{Reinhold Egger and Hermann Grabert}
\address{Fakult\"at f\"ur Physik, Albert-Ludwigs-Universit\"at,
Hermann-Herder-Stra{\ss}e 3, D-79104 Freiburg, Germany }
\maketitle
\widetext
\begin{abstract}
Screening in one-dimensional metals is studied for arbitrary 
electron-electron interactions. It is shown that for
finite-range interactions (Luttinger liquid) electroneutrality
is violated.  This apparent inconsistency can be traced to the presence of
external screening gates responsible for the effectively 
short-ranged Coulomb interactions.  We also draw attention to the
breakdown of linear screening for wavevectors close to $2k_F$.
\end{abstract}

\narrowtext

Screening is one of the most important and useful concepts in 
condensed-matter physics \cite{debye,pines}. 
If some external charge is brought into a conductor,
the internal charge carriers will reorganize with the new
distribution of charge eliminating the electric field at large
distances.  The  screened potential
set up by the external charge together with its screening  
cloud is then rather short-ranged.
Typically, one ends up with only weakly correlated systems,
despite of originally long-range electrostatic forces.

Most theories employ linear screening as working hypothesis,
where the effects of  (possibly time-dependent) external test charges
onto the conduction electrons
are determined by linear response theory.
Given the validity of linear screening,
 the wavevector- and frequency-dependent
dielectric function $\epsilon(q,\omega)$ contains all relevant
information about screening.  
An important result of the theory is the Friedel sum rule \cite{friedel},
which states that the total electronic screening charge 
exactly compensates any external (impurity) charge brought 
into the system.
This charge neutrality requirement on large scales
arises because in equilibrium there can be
no net electric field at large distances.  
The validity of the Friedel sum rule is usually
taken for granted, generally by referring to the 
analysis in Ref.\cite{langer} where this was proven explicitly for the
Anderson model. 

In this paper, we discuss screening and the
Friedel sum rule for interacting electrons in one dimension (1D).
At low energy scales, provided no lattice or spin instabilities
are present, the properties of 1D fermions can be described
by the Luttinger liquid model \cite{schulz} 
(or slight generalizations thereof, see below).
The Luttinger liquid is a strongly correlated 1D metal which does  
not support the existence of Landau quasi-particles.
It is of importance for a number of applications of current interest,
e.g., quantum wires in semiconductor heterostructures in the
limit of one transport channel  \cite{gogrev}, transport in
carbon nanotubes \cite{tube}, or quasi-1D organic conductors \cite{voit},
to mention a few. Here we show that in a Luttinger liquid the 
screening charge does not balance the impurity charge.
\break{1.2in}
Nevertheless, Friedel's phase shift sum rule
\cite{friedel} remains valid in terms of an adequately defined
phase shift.

Our investigation of screening in 1D
is based on the standard bosonization method \cite{gogolin}.
From a comparison with alternative techniques,
this method is known to give a proper description of 1D fermions
at low energy scales (we consider zero temperature below).
Since spin and charge are decoupled in a Luttinger liquid, 
it is sufficient to study only the spinless case in the following,
with the same conclusions applying to spin-$\frac12$ electrons.
The low-lying excitations in a spinless Luttinger liquid
are described by a bosonic phase field $\theta(x)$, 
in terms of which the electron density operator can be written in the form
\begin{equation} \label{dens}
\rho(x) = \frac{k_F}{\pi} + \frac{1}{\sqrt{\pi}} \partial_x \theta(x)
+ \frac{k_F}{\pi} \cos [ 2k_F x + \sqrt{4\pi} \theta(x) ] \;.
\end{equation}
The first term describes the mean charge density (which is supposedly
neutralized by a positive background), 
the second term gives 
long-wavelength ($q\simeq 0$) fluctuations, and the last term
yields rapidly oscillating $(|q|\simeq 2k_F$) contributions.
Putting $\hbar=1$, the Luttinger liquid Hamiltonian is then given by 
\cite{schulz,gogolin}
\begin{eqnarray}  \label{h0}
H_0 &=& \frac{v_F}{2} \int dx \,[\Pi^2(x) + (\partial_x \theta)^2 ]
\\ \nonumber
&+& \frac{1}{2\pi}\int dx dx' \, \partial_x \theta(x) U(x-x') \partial_{x'}
\theta(x') \;,
\end{eqnarray}
where $\Pi(x)$ is the canonically conjugate momentum to $\theta(x)$
and $v_F$ the Fermi velocity.
The potential $U(x)$  can describe either an unscreened
Coulomb interaction,  $U(x)\sim 1/|x|$,
or an externally screened finite-range potential which arises due to the
presence of mobile charge carriers close to the 1D metal, e.g., on 
screening gates or other nearby chains.  To simplify notation, we shall make 
the inessential assumption that $U(x)$ is sufficiently long-ranged
 such that its Fourier transform
 $\widetilde{U}(q)$ has a very small component at $q = 2 k_F$. Then
electron-electron backscattering can be neglected, 
as implied in Eq.~(\ref{h0}).
Even if this should not be the case, the bosonization technique
can still be applied and yields qualitatively the same results. 
The Luttinger liquid model in the strict sense is 
obtained by effectively using  a 
local interaction $U(x)= \widetilde{U}(0) \delta(x)$.
Here we shall employ the usual dimensionless
Luttinger liquid interaction parameter $g$  defined as
\begin{equation}\label{gdef}
    g = [1+\widetilde{U}(0)/\pi v_F]^{-1/2}
\end{equation}
for all finite-range interactions. Then $g=1$ is the non-interacting
limit, while for repulsive interactions we have $g<1$. 
In the absence of screening gates, $g$ approaches zero in an infinitely
long system \cite{schulz,gogolin}.

To study  screening properties, we now consider some
external time-dependent charge distribution $eQ(x,t)$ brought into the 
system. The interaction with the 1D metal reads
\begin{equation} \label{ext}
H_Q (t) = \int dx dx'\, Q(x,t) U(x-x') \rho(x')  \;.
\end{equation} 
In view of the representation (\ref{dens}) for the electronic
density, there are two  contributions.
The first comes from the $q\simeq 0$
component, and the second from the $q\simeq 2k_F$ part. 
We note that the interaction potentials 
in Eqs.~(\ref{h0}) and (\ref{ext}) are the same because
we deal with the internally unscreened, microscopic
interaction at this stage.

Let us first discuss the long-wavelength ($|q|\ll 2k_F$) response
of the electrons. Ignoring the $2k_F$ part in Eq.~(\ref{ext}), the
now Gaussian Hamiltonian yields straightforwardly
\begin{equation} \label{slow}
\langle \rho(q,\omega) \rangle = 
\frac{v_F}{\pi} \frac{q^2 \widetilde{U}(q)}{
\omega^2 - \omega^2(q) } Q(q,\omega) 
\end{equation}
with the plasmon dispersion relation
\[
\omega(q) = v_F |q| \sqrt{1+\widetilde{U}(q)/\pi v_F} \;.
\]
Apparently, in the long-wavelength limit, 
the Luttinger liquid model implies linear screening
\[
\langle \rho(q,\omega)\rangle = \widetilde{U}(q) \chi(q,\omega) Q(q,\omega)
\]
with the polarizability $\chi(q,\omega)$.
The response of the electrons to  
$Q(x,t)$ is thus fully described by a dielectric function.
One finds from Eq.~(\ref{slow}) and the definition \cite{pines}
\begin{equation}\label{defeps}
\epsilon^{-1}(q,\omega) = 1+ \widetilde{U}(q) \chi( q,\omega) 
\end{equation}
the small-$q$ result
\[
\epsilon^{-1}(q,\omega) = 1+ \frac{v_F}{\pi} \frac{q^2 
\widetilde{U}(q)}{\omega^2-\omega^2(q)} \;.
\]
In the static case, $\omega=0$, this yields
\begin{equation}\label{static}
\epsilon(q) = 1+ \widetilde{U}(q)/\pi v_F\;.
\end{equation}
We mention in passing that for large impurity charge $eQ$ 
the bosonization approach breaks down \cite{mahan}. For instance,
Eq.~(\ref{slow}) would incorrectly predict that the 
electron density becomes negative
for sufficiently large $Q$.

One can now define the internally screened interaction potential
$\widetilde{U}_{\rm eff}(q) = 
\widetilde{U}(q)/\epsilon(q)$ which determines the effective
 potential between two charges \cite{pines}.
From Eq.~(\ref{static}), its long-wavelength part is
$\widetilde{U}_{\rm eff}(q) = \widetilde{U}(q)/[1+\widetilde{U}(q)/\pi v_F]$,
which gives for a finite-range interaction
 $\widetilde{U}_{\rm eff}(q) = g^2 \widetilde{U}(q)$ as $q\to 0$. For an
externally unscreened $1/|x|$ interaction, one has $\widetilde{U}(q)= 2e^2
|\ln(qd)|$, where $d$ is the width of the 1D channel \cite{schulz}.
Apart from a hard core at small distances, this
leads to the large-distance behavior valid at $|x|\gg d$
\begin{equation}
U_{\rm eff}(x) \sim \frac{1}{|x| \ln|x/d|} \;.
\end{equation}
Therefore the long-range character
of the interaction is not significantly reduced. The only 
logarithmic suppression of the $1/|x|$ law explicitly
demonstrates the very weak screening in 1D.

The condition for perfect screening 
\[
 \epsilon^{-1}(q\to 0,\omega=0) \to 0
\]
is seen to be violated in any finite-range model.
This follows directly from Eq.~(\ref{static})
since $\epsilon(q\to 0) = 1/g^2$. The implications
are best discussed for a point charge sitting at $x=0$, i.e.,
 $Q(x,t)=Q \delta(x)$. The corresponding 
long-wavelength response is given in Eq.~(\ref{slow}).
For the total screening charge, $eQ_s=e\int dx \langle \rho(x)
 \rangle$, this leads to the strikingly simple result
\begin{equation}\label{sum}
Q_s =  - (1-g^2) Q \;,
\end{equation}
where $g$ has been defined in Eq.~(\ref{gdef}).
We stress that this relation holds for any finite-range Coulomb 
interaction.
Asserting that the $2k_F$ Friedel oscillation in the 
charge density does not contribute to the total screening charge
 (see below), 
we observe that only a fraction $1-g^2 < 1$ of the
external charge $Q$ is screened by the conduction electrons.
Therefore the {\em electroneutrality condition} for
impurity plus screening charge, $Q_s+Q=0$, is apparently
{\em violated}\, in models with a finite-range interaction.
Of course, this reasoning carries over to lattice models
with effectively short-ranged interactions, e.g., the 1D Hubbard model.
For a long-range $1/|x|$ interaction, the parameter $g$ 
effectively approaches zero, and electroneutrality is then seen 
to hold.

The result (\ref{sum}) can also be obtained by a phase shift
consideration.  Forward scattering due to a point-like 
impurity charge $Q \delta(x)$ in Eq.~(\ref{ext}) can be
eliminated by the standard unitary transformation
\[
  U = \exp\{ - i\sqrt{\pi} \int dx \alpha(x) \phi(x) \} \;,
\]
where $\phi(x)$ is the dual field to $\theta(x)$ \cite{gogolin},
and the Fourier transform of $\alpha(x)$ is
\begin{equation}\label{alpha}
\widetilde{\alpha}(q) = - \frac{\widetilde{U}(q)/\pi v_F}
{1+ \widetilde{U}(q)/\pi v_F} \, Q \;.
\end{equation}
Comparing with Eq.~(\ref{slow}) for $\omega=0$,
 the induced electronic density is
simply $\langle \rho(x) \rangle = \alpha(x)$.
The unitary transformation $U$ now leads to   a
phase shift appearing, e.g., in the $2k_F$ part of Eq.~(\ref{dens}),
which takes the form
\[
\eta(x) =  \pi \int dx' \, {\rm sgn}(x-x') \alpha(x') \;.
\]
Defining the asymptotic phase shift $\eta_F = \eta(x\to\infty)$,
we find
\begin{equation} \label{fried}
 Q_s =  \int dx \, \alpha(x) = \eta_F/\pi \;.
\end{equation}
Despite the apparent violation of electroneutrality, 
Friedel's phase shift sum rule \cite{friedel}
is seen to hold.   Clearly, the phase shift $\eta_F$ 
characterizes some screened impurity charge and not the bare charge $Q$ 
brought into the system.  Finally, using Eqs.~(\ref{alpha}) 
and (\ref{fried}), one may verify Eq.~(\ref{sum}) again.

We mention that the conventional Fermi liquid case is not 
directly included in Eq.~(\ref{sum}) as the simple limit $g\to 1$. 
For a Fermi liquid, one assumes that quasi-particles with
 good screening exist and then adds a local potential scatterer
in order to derive the Friedel sum rule \cite{friedel}.
Its scattering strength is related to a
phase shift $\eta_F$, and the screening charge is
 $Q_s=\eta_F/\pi$ as in Eq.~(\ref{fried}).
By interpreting $\eta_F/\pi$ as the impurity charge,
the Friedel sum rule is then in fact
{\em imposed}\, as a consistency relation ensuring electroneutrality
of the system.
In contrast, putting $g=1$ for the Luttinger liquid model 
would imply a non-interacting system (Fermi gas) rather than a Fermi liquid.

The physical reason for the apparent failure of
electroneutrality in finite-range models  
are induced charges outside the 1D system,
e.g., on external screening gates, which 
cause the finite range of the interaction.
These other conductors also contribute to the total 
screening charge.  To give a concrete example, consider
the gated 1D quantum wire shown in Fig.~1. For a wire of width $d$,
the presence of a two-dimensional gate at a distance $D$
leads to a short-range interaction characterized by 
\[
g = \left\{ 1+ \frac{2e^2}{\pi v_F} \ln (2D/d) \right\}^{-1/2} \;.
\]
The induced 1D charge density [integrated over the 
$y$-direction] on the gate, $e\rho_G(x)$,  is obtained as
\[
\rho_G(x) = - \frac{D}{\pi} \int dx' \,\frac{q(x')}{D^2+(x-x')^2} \;,
\]
where $q(x')$ is the total density in the wire
(including the impurity).  Integration over $x$
gives straightforwardly the total induced charge on the gate
\[
Q_G = \int dx\, \rho_G(x) = -(Q + Q_s) \;.
\]
In effect, the electroneutrality condition in the form
\begin{equation}  \label{total}
Q_s+Q+Q_G=0
\end{equation}
is then restored for the total system including the
screening gates. Ignoring the screening 
gates implicitly used to derive the Luttinger liquid model
is thus responsible for the modified condition (\ref{sum}) 
within the 1D system.  Parenthetically, we note that if
the charge $Q$ is not put directly into the 1D system
but some distance away, the screening charge $Q_s$ is not
given by Eq.~(\ref{sum}) anymore, yet Eq.~(\ref{total})
will still hold.

\begin{figure}
\epsfysize=4cm
\epsffile{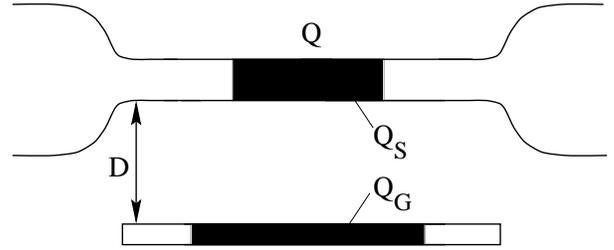}
\caption[]{\label{fig1} 
Schematic view of a 1D quantum wire with short-range Coulomb 
interaction due to the
presence of a 2D screening gate located a distance $D$ away from the
wire. The bare impurity charge is $Q$, the
direct screening charge within the 1D system is $Q_s=-(1-g^2) Q$
[see Eq.~(\ref{sum})], and the induced charge on the 
screening gate is $Q_G$.}
\end{figure}

The underscreening of an external charge brought into
the 1D system should also be experimentally observable.
If charge is injected into, e.g.,
a carbon nanotube constituting a perfect experimental
realization of a 1D conductor \cite{tube},
the resulting image charge on nearby external screening gates
can be detected by capacitance spectroscopy \cite{ashoori}
or by using highly sensitive single-electron transistor (SET)
electrometers on top of a scanning probe tip \cite{set}.

So far we have discussed the long-wavelength part of
the electronic response only. Turning now to the $2k_F$ part \cite{foot0}
and assuming linear screening, we have to 
compute the polarizability $\chi$, which is essentially
the double-Fourier transformed density-density correlation 
function of the unperturbed conductor
\begin{equation}\label{corr}
\chi(q, i \omega) = - \int dx d\tau \, 
e^{-i\omega \tau - i qx} 
\langle T_\tau \rho(x,\tau) \rho(0,0) \rangle \;.
\end{equation}
Here $T_\tau$ is the time-ordering operator in Euclidean time,
 and one has to  analytically continue Eq.~(\ref{corr})
to real frequencies, $i\omega \to \omega+i0^+$, in order
to obtain $\chi(q,\omega)$ needed in Eq.~(\ref{defeps}).
The $2k_F$ part of $\chi$ for a Luttinger liquid is found to read
\begin{eqnarray*} 
\chi(q,i\omega) & =& - \frac{C_g}{\pi v_F} \sum_{p,s=\pm}
\left( \frac{is\omega}{v_F k_F} + \left| \frac{q}{k_F} -2p \right|
\right)^{2g-2}\\ &\times&
F \left(2-2g,1-g;2-g;\frac{is\omega-v_F|q-2pk_F|}
{is\omega+v_F|q-2pk_F|} \right)\;,
\end{eqnarray*}
where $F$ denotes the hypergeometric function and
$C_g= 4^{-g} \Gamma(2-2g)/\Gamma(g) \Gamma(2-g)$.
In the static case, this gives algebraic singularities 
\begin{equation}\label{sing}
\chi(q) = - \frac{\bar{C}_g} {\pi v_F} \sum_{p=\pm} \left|\frac{q}{k_F}
-2p\right|^{2g-2} \;,
\end{equation}
with the numerical constant
\[ 
\bar{C}_g= \frac{\sqrt{\pi} \Gamma(2-2g)}{2\Gamma(g)\Gamma(3/2-g)} \;.
\]
From these algebraic singularities one would infer a Friedel oscillation
 decaying as $\langle\rho(x)\rangle \sim \cos(2k_F x) x^{1-2g}$
and a similar contribution to  the screened interaction potential 
$U_{\rm eff}(x)$. However, this represents only the first order in the
perturbation expansion for the Friedel oscillation and determines
merely the short-distance behavior, while
the long-distance behavior of the Friedel oscillation 
necessitates a calculation in all orders of the impurity strength 
\cite{egger,leclair}.  An important implication is the 
{\em breakdown of linear screening} for the $2k_F$ electronic response.
This breakdown occurs for arbitrarily small impurity charge $eQ$ 
at wavevectors close enough to $2k_F$.
Following the results of Ref.\cite{egger}, the
singularity exponent $2g-2$ for $q\to 2k_F$ in Eq.~(\ref{sing})
is turned into $g-1$. As a consequence, the
effective screened potential as well as the Friedel 
oscillation asymptotically decay as $\sim \cos(2k_F x) x^{-g}$.
Due to the intrinsically nonlinear screening, 
the dielectric function is of rather limited use 
for wavevectors close to $2k_F$.

The bosonization approach naturally separates the density operator
(\ref{dens}), and therefore also the electronic screening response,
into a slow and a fast $2k_F$ part.  
The total screening charge is determined
by the $q=0$ component of the induced charge density,
which in turn is exclusively given by the slow part (\ref{slow}).
Therefore the Friedel oscillation obtained from the bosonized 
$2k_F$ part of Eq.~(\ref{dens}) does not contribute 
to the total screening charge.
In a microscopic calculation, one will in general not be able to 
separate the slow and the fast components so nicely, but within 
the bosonization approach, a quite simple derivation of the
total screening charge (\ref{sum}) is possible.

To conclude, we have investigated screening in one dimension.
We have shown that electroneutrality is not obeyed in models with a
finite-range (screened) Coulomb interaction. In a 1D metal, the
total induced screening charge is given by $Q_s = -(1-g^2) Q$,
where $Q$ is the impurity charge and $g$ the Luttinger liquid 
interaction parameter.
To resolve this apparent inconsistency, one needs
to take into account induced charges on screening gates.
 
We wish to thank H. Schoeller and J. Voit for valuable discussions.
Support by the Deutsche Forschungsgemeinschaft (Bonn) is 
acknowledged.


\begin{references}
\bibitem{foot0} Note that there is no $2 k_F$ Friedel oscillation 
contribution built into Eq.~(\ref{slow})
although translational invariance is broken by the 
presence of the impurity charge.
\end{references}
\end{document}